\documentclass[11pt,superscriptaddress,aps,prd,preprint]{revtex4}
\usepackage[dvips]{graphicx}
\usepackage{amsfonts}
\newcommand{\be}{\begin{equation}}
\newcommand{\en}{\end{equation}}
\newcommand{\ba}{\begin{eqnarray}}
\newcommand{\ea}{\end{eqnarray}}

\newcommand{\Slash}[1]{{#1}\!\!\!/}
\newcommand{\RM}[1]{\mathrm{#1}}

\begin{document}

\title{On the ambiguities in the effective action in Lorentz-violating gravity}

\author{M. Gomes}
\affiliation{Instituto de F\'{\i}sica, Universidade de S\~ao Paulo\\
Caixa Postal 66318, 05315-970, S\~ao Paulo, SP, Brazil}
\email{mgomes,ajsilva,tmariz,jroberto@fma.if.usp.br}

\author{T. Mariz}
\affiliation{Instituto de F\'{\i}sica, Universidade de S\~ao Paulo\\
Caixa Postal 66318, 05315-970, S\~ao Paulo, SP, Brazil}
\email{mgomes,ajsilva,tmariz,jroberto@fma.if.usp.br}

\author{J. R. Nascimento}
\affiliation{Departamento de F\'{\i}sica, Universidade Federal da Para\'{\i}ba\\
Caixa Postal 5008, 58051-970, Jo\~ao Pessoa, Para\'{\i}ba, Brazil}
\email{mgomes,ajsilva,tmariz,jroberto@fma.if.usp.br}

\author{E. Passos}
\affiliation{Departamento de F\'{\i}sica, Universidade Federal da Para\'{\i}ba\\
Caixa Postal 5008, 58051-970, Jo\~ao Pessoa, Para\'{\i}ba, Brazil}
\email{mgomes,ajsilva,tmariz,jroberto@fma.if.usp.br}

\author{A. Yu. Petrov}
\affiliation{Departamento de F\'{\i}sica, Universidade Federal da Para\'{\i}ba\\
Caixa Postal 5008, 58051-970, Jo\~ao Pessoa, Para\'{\i}ba, Brazil}
\email{mgomes,ajsilva,tmariz,jroberto@fma.if.usp.br}

\author{A. J. da Silva}
\affiliation{Instituto de F\'{\i}sica, Universidade de S\~ao Paulo\\
Caixa Postal 66318, 05315-970, S\~ao Paulo, SP, Brazil}
\email{mgomes,ajsilva,tmariz,jroberto@fma.if.usp.br}

\begin{abstract}
We investigate the occurrence of ambiguities for Lorentz-violating gravitational Chern-Simons term. It turns out that this term is accompanied by a coefficient depending on an undetermined parameter, due to an arbitrariness in the choice of the conserved current.
\end{abstract}

\maketitle

\section{Introduction}

The possibility of Lorentz symmetry breaking which is being
intensively discussed nowadays, opens a wide range for new physical
phenomena~\cite{Kostel}. Besides the generation of new Lorentz-breaking
couplings, one important consequence of that investigation is
the ambiguity found in the finite contributions to the effective
action~\cite{Jackiw}. Recently, this ambiguity was shown to appear in
a very specific form, that is, it manifests in a finite but otherwise
undetermined extra term in the one-loop effective action of the 
Lorentz-breaking electrodynamics within the functional integral
formalism~\cite{Chung}. However, as in the framework of the
functional integral method there are different approaches for
calculating the one-loop effective action, a natural
question is whether such extra term arises within different schemes
of the one-loop calculations. 

Another natural problem is the study of the possible Lorentz-breaking
terms in gravity models. In this context, the gravitational Chern-Simons term,
firstly introduced in~\cite{Jack1}, was shown to arise as a quantum
correction in the theory of a spinor field coupled to gravity in a
Lorentz-breaking manner, both in the linearized approximation \cite{Mar2} as well as in the
full-fledged theory \cite{ptime}. Further, a number of physical
aspects of this term, such as, properties of the energy-momentum
tensor and other conserved currents, and the possible black hole
solutions, were studied \cite{aspects}.
 
In this paper, we carry out the calculation of the gravitational
Chern-Simons term using functional integral approach similarly to
\cite{Chung}, but securing that the gravitational gauge invariance  
is respected. 

The structure of the paper looks like follows. In the section 2, a contribution to the effective action generted by the modification of an integral measure is discussed. The section 3 is devoted to obtaining of the perturbative contribution to the Chern-Simons action. In the Summary, the results are discussed.

\section{Gravitational Chern-Simons action}

Let us, following \cite{Chung}, describe the functional
integral approach for the theory involving spinors coupled to gravity
in a Lorentz-breaking manner. The action in which we are interested includes a
Lorentz-breaking term proportional to a constant vector $b^\mu$ \cite{Mar2},
\be
S = \int d^4x\,e\,\left(\frac i2{e^\mu}_a\bar\psi\gamma^a\!\!\stackrel{\leftrightarrow}{D}_\mu\!\!\psi - \,m\bar\psi\psi  - b_{\mu}j^{\mu}_5\right),
\en
where $e^{\mu}_{\phantom{a}a}$ is the tetrad (vierbein) and $e\equiv
{\rm det}\,e^{\mu}_{\phantom{a}a}$. The covariant derivative is given by
\be
\label{D}
D_{\mu}\psi=\partial_{\mu}\psi-i\omega_{\mu}\psi,
\en
where $\omega_\mu=\frac{1}{4}\omega_{\mu bc}\sigma^{bc}$ is the spin
connection and $\sigma^{bc}=\frac{i}{2}[\gamma^b,\gamma^c]$. The current is chosen as 
\be
\label{curr}
j^{\mu}_5={e^\mu}_a\bar{\psi}\gamma^a\gamma_5\psi+K^{\mu},
\en
with a convenient definition for $K^{\mu}$ to be specified later. 

Within this work we consider the gravitational field to be purely external.
The generating functional for this theory looks like
\ba
Z=\int {\cal D}\psi{\cal D}\bar{\psi}\exp(iS).
\ea 
Under the
chiral transformations of the spinor field
\ba
\label{tra}
\psi(x)&\to&\exp[i\alpha(x)\gamma_5]\psi(x),\nonumber\\
\bar{\psi}(x)&\to&\bar{\psi}(x)\exp[i\alpha(x)\gamma_5],
\ea
the integral measure $D\psi D\bar{\psi}$ is corrected 
by the multiplier \cite{Fuji1}
\ba
\Delta=\exp\left\{-2i\int d^4x\,\alpha(x)\lim_{M\to\infty}{\rm
    Tr}\left[\gamma_5e^{-\left(e\,{e^\mu}_a\gamma^a D_\mu/M
\right)^2}\right]\right\},
\ea
where the covariant derivative is given by (\ref{D}). Proceeding as in \cite{Fuji,Fuji1}, we arrive at
\ba
\Delta&=&
\exp\left(-\frac{i}{384\pi^2}\int
  d^4x\,\alpha(x)\,\epsilon^{\mu\nu\lambda\rho}
R_{\mu\nu ab}{R_{\lambda\rho}}^{ab}
\right).
\ea

It is easy to see (cf. \cite{Jack1}) that 
\ba
\frac{1}{4}\epsilon^{\mu\nu\lambda\rho}R_{\mu\nu ab}{R_{\lambda\rho}}^{ab}=\partial_{\mu}L^{\mu},
\ea 
where
\ba
L^{\mu}=\epsilon^{\mu\nu\lambda\rho}\left(
\omega_{\nu ab}\partial_\lambda{\omega_\rho}^{ba}-
\frac{2}{3}\omega_{\nu ab}{\omega_{\lambda}}^{bc}
{\omega_{\rho c}}^a\right)
\ea
is the Chern-Simons current.

By choosing $\alpha(x)=-x_\mu b^\mu$ and integrating by parts, we then find
\ba
\Delta&=&\exp\left(-\frac{i}{96\pi^2}\int d^4x\,b_{\mu}\epsilon^{\mu\nu\lambda\rho}\left[\omega_{\nu ab}\partial_\lambda{\omega_\rho}^{ba}-\frac{2}{3}\omega_{\nu ab}{\omega_{\lambda}}^{bc}
{\omega_{\rho c}}^a\right]
\right).
\ea
The argument of the exponential exactly reproduces the form of the Chern-Simons term from \cite{Jack1,Mar2,ptime} (see also \cite{Fuji} for a general discussion of gravitational anomalies).

Under the transformations (\ref{tra}) the action is changed to
\ba
S = \int d^4x\,e\left[\frac i2{e^\mu}_a\bar\psi\gamma^a\!\!\stackrel{\leftrightarrow}{D}_\mu\!\!\psi -{e^\mu}_a(\partial_{\mu}\alpha)\bar{\psi}
\gamma^a\gamma_5\psi- m\bar\psi e^{2i\alpha\gamma_5}\psi - 
b_{\mu}j^{\mu}_5\right].
\ea
Let us now obtain the explicit form of the current $j^{\mu}_5$ introduced in
Eq. (\ref{curr}). It is well known that its first term, that is,
${e^\mu}_a\bar{\psi}\gamma^a\gamma_5\psi$, is gauge invariant. At the same
time, the requirement of gauge invariance of the action imposes the restriction that the integral of $K^{\mu}$
must be gauge invariant, although $K^{\mu}$ itself is not necessarily
invariant. It is easy to verify (details of the discussion can
be found in \cite{Jack1}), that, in the case of gravity, the
acceptable form of this vector is the Chern-Simons 
topological current
\ba
e\,K^{\mu}=-C\,\epsilon^{\mu\nu\lambda\rho}\left(
\omega_{\nu ab}\partial_\lambda{\omega_\rho}^{ba}-
\frac{2}{3}\omega_{\nu ab}{\omega_{\lambda}}^{bc}
{\omega_{\rho c}}^a\right),
\ea
where $C$ is some constant and $\epsilon^{\mu\nu\lambda\rho}=e\,{e^\mu}_a{e^\nu}_b{e^\lambda}_c{e^\rho}_d\epsilon^{abcd}$, which evidently will give rise to a gravitational Chern-Simons term added in the action. As a result, the generating functional of the theory, in which the modified measure and the additional term generated by the conserved current are taken into account, can be written as
\ba
Z&=&\exp\left(-\frac i{96\pi^2} \int d^4x\,b_\mu L^\mu\right)\times\exp\left(i C \int d^4x\,b_\mu L^\mu\right)\\&&\times
\int D\psi D\bar{\psi}\exp\left[i\int d^4x
  \,e\left(\frac i2{e^\mu}_a\bar\psi\gamma^a\!\!\stackrel{\leftrightarrow}{D}_\mu\!\!\psi -  m\bar\psi
    e^{2i\,x\cdot b\,\gamma_5}\psi \right)\right].\nonumber
\ea
By performing the fermionic integration, $Z = e^{i \Gamma_\RM{eff}[\omega]}$, the whole one-loop effective action $\Gamma_\RM{eff}[\omega]$ of the gravitational field is given by
\ba
\label{total}
\Gamma_\RM{eff}[\omega]=\left(C-\frac{1}{96\pi^2}\right)\int d^4x
b_{\mu}\,\epsilon^{\mu\nu\lambda\rho}\left(
\omega_{\nu ab}\partial_\lambda{\omega_\rho}^{ba}-
\frac{2}{3}\omega_{\nu ab}{\omega_{\lambda}}^{bc}
{\omega_{\rho c}}^a\right)+S'_\RM{eff}[\omega],
\ea
with
\be
S'_\RM{eff}[\omega] = -i\RM{Tr}\ln\left(\frac i2e\,{e^\mu}_a\gamma^a\!\!\stackrel{\leftrightarrow}{D}_\mu -  \,e\,m e^{2i\,x\cdot b\,\gamma_5}\right),
\en
where $\RM{Tr}$ stands for the trace over Dirac matrices as well as trace over the integration in momentum and coordinate spaces.

\section{Perturbative induction of the Chern-Simons action}

To complete the calculation of the one-loop gravitational Chern-Simons action, we must compute $S'_\RM{eff}[\omega]$ up to first order in the Lorentz-breaking vector $b_\mu$ which is expressed as
\be\label{Seffw}
S'_\RM{eff}[\omega] = -i\RM{Tr}\ln\left(\frac i2e\,{e^\mu}_a\gamma^a\!\!\stackrel{\leftrightarrow}{\partial}_\mu - \,m - \,2ie\,m\,x\!\cdot\!b\,\gamma_5 + e\,{e^\mu}_a\gamma^a\omega_\mu \right).
\en
Here we use the weak field approximation in which the vierbein and the
connection are expressed in terms of the metric fluctuation $h_{\mu\nu}$ (which is
the only dynamical field in the weak field approximation of gravity) as $e_{\mu a}=\eta_{\mu a}+\frac{1}{2}h_{\mu a}$ and 
$\omega_{\mu ab}=\frac{1}{2}(\partial_b h_{\mu a}-\partial_a h_{\mu b})$, thus we have
\be\label{Seffh}
S'_\RM{eff}[h] = -i\RM{Tr}\ln\left[i\Slash{\partial} - m - 2im\,x\!\cdot\!b\,\gamma_5 - \frac i4 h_{\mu\nu}\gamma^\mu\!\!\!\!\stackrel{\leftrightarrow}{\,\,\partial\,^\mu}  - \frac i{16}(h_{\mu\alpha}\!\!\!\stackrel{\leftrightarrow}{\,\,\partial_\lambda}\!{h_\nu}^\alpha)\Gamma^{\mu\nu\lambda}\right],
\en
where $\Gamma^{\mu\nu\lambda}$ in the antisymmetrized product of three Dirac matrices, i.e., $\Gamma^{\mu\nu\lambda}=\frac16\left(\gamma^\mu\gamma^\nu\gamma^\lambda - \gamma^\mu\gamma^\lambda\gamma^\nu + \gamma^\nu\gamma^\lambda\gamma^\mu - \gamma^\nu\gamma^\mu\gamma^\lambda + \gamma^\lambda\gamma^\mu\gamma^\nu - \gamma^\lambda\gamma^\nu\gamma^\mu\right)$. Up to the field independent factor $-i\RM{Tr}\ln\left(i\Slash{\partial} - m - 2im\,x\!\cdot\!b\,\gamma_5\right)$, which may be absorbed in the normalization of the generating functional, we rewrite the Eq. (\ref{Seffh}) as
\be\label{Seffh2}
S_\RM{eff}^{(n)}[h] = i\RM{Tr}\sum_{n=1}^\infty\frac 1n \left\{\frac 1{i\Slash{\partial} - m - 2im\,x\!\cdot\!b\,\gamma_5} \left[\frac i4 h_{\mu\nu}\gamma^\mu\!\!\!\!\stackrel{\leftrightarrow}{\,\,\partial\,^\mu}  + \frac i{16}(h_{\mu\alpha}\!\!\!\stackrel{\leftrightarrow}{\,\,\partial_\lambda}\!{h_\nu}^\alpha)\Gamma^{\mu\nu\lambda}\right]\right\}^n.
\en

As we are interested in the radiative induction of the gravitational Chern-Simons action, here and in what follows we restrict ourselves to the second order in $h_{\mu\nu}$ and first order in $b_\mu$. Firstly, let us analyze the terms that come from $n=1$, as follows
\be
S_\RM{CS}^{(1)}[h] = i\RM{Tr}\frac 1{i\Slash{\partial} - m}2im\,x\!\cdot\!b\,\gamma_5\frac 1{i\Slash{\partial} - m} \,\frac i{16}(h_{\mu\alpha}\!\!\!\stackrel{\leftrightarrow}{\,\,\partial_\lambda}\!{h_\nu}^\alpha)\Gamma^{\mu\nu\lambda}.
\en
In order to carry out the traces over the integration in spaces we must use the prescription $i\partial_\mu\to p_\mu$ and $x_\mu\to i\frac{\partial}{\partial p_\mu}$ \cite{Ait,Cha,Che,Zuk}, which is more convenient due to the presence of $\alpha(x)=-x\!\cdot\!b$, so that we get
\be
S_\RM{CS}^{(1)}[h] = i\,\RM{tr}\int d^4x \int \frac{d^4p}{(2\pi)^4}S(p)(-2m)\frac{\partial}{\partial p^\beta}b^\beta\gamma_5S(p) \,\frac i{16}(h_{\mu\alpha}\!\!\!\stackrel{\leftrightarrow}{\,\,\partial_\lambda}\!{h_\nu}^\alpha)\Gamma^{\mu\nu\lambda},
\en
where the symbol $\RM{tr}$ means that the trace is only over Dirac matrices and $S(p)=(\Slash{p}-m)^{-1}$. Now, by using the identity
\be
\frac{\partial}{\partial p^\beta} S(p) = S(p)\gamma_\beta S(p),
\en
we have
\ba
S_\RM{CS}^{(1)}[h] &=& i\int d^4x\,h_{\mu\nu}\,\Pi_a^{\mu\nu\alpha\beta}\,h_{\alpha\beta},
\ea
where
\be\label{inta}
\Pi_a^{\mu\nu\alpha\beta} = -\frac m4 \RM{tr} \int\frac{d^4p}{(2\pi)^4} S(p)\gamma_5S(p)\Slash{b}S(p)\Gamma^{\mu\alpha\lambda}\eta^{\nu\beta}\partial_\lambda.
\en

Finally, let us single out the terms coming from $n=2$, given by
\ba
S_\RM{CS}^{(2)}[h] &=& \frac i2 \RM{Tr}\frac 1{i\Slash{\partial} - m}2im\,x\!\cdot\!b\,\gamma_5\frac 1{i\Slash{\partial} - m} \frac i4 h_{\mu\nu}\gamma^\mu\!\!\!\!\stackrel{\leftrightarrow}{\,\,\partial\,^\mu}\!\! \frac 1{i\Slash{\partial} - m}\frac i4 h_{\alpha\beta}\gamma^\alpha\!\!\!\!\stackrel{\leftrightarrow}{\,\,\partial\,^\beta}\!\! \nonumber\\
&& + \frac i2 \RM{Tr}\frac 1{i\Slash{\partial} - m} \frac i4 h_{\mu\nu}\gamma^\mu\!\!\!\!\stackrel{\leftrightarrow}{\,\,\partial\,^\mu}\!\! \frac 1{i\Slash{\partial} - m}2im\,x\!\cdot\!b\,\gamma_5\frac 1{i\Slash{\partial} - m} \frac i4 h_{\alpha\beta}\gamma^\alpha\!\!\!\!\stackrel{\leftrightarrow}{\,\,\partial\,^\beta}\!\!. 
\ea
Besides the above prescription, we also use the identity \cite{Ait,Cha,Che,Zuk}
\be
h_{\mu\nu}(x)S(p)=S(p-i\partial)h_{\mu\nu}(x)
\en
in order to disentangle the traces over $x_\mu$ and $p_\mu$, and thus we arrive at
\ba
S_\RM{CS}^{(2)}[h] &=& \frac i2\int d^4x \,h_{\mu\nu}\left(\Pi_b^{\mu\nu\alpha\beta}+\Pi_c^{\mu\nu\alpha\beta}\right)h_{\alpha\beta},
\ea
where
\be\label{intb}
\Pi_b^{\mu\nu\alpha\beta} = \RM{tr} \int\frac{d^4p}{(2\pi)^4} S(p)(-2m)\frac{\partial}{\partial p^\rho}b^\rho\gamma_5S(p)\frac14(2p^\mu-i\partial^\mu)\gamma^\nu S(p-i\partial)\frac14(2p^\alpha-i\partial^\alpha)\gamma^\beta
\en
and
\be\label{intc}
\Pi_c^{\mu\nu\alpha\beta} = \RM{tr} \int\frac{d^4p}{(2\pi)^4} S(p)\frac14(2p^\mu-i\partial^\mu)\gamma^\nu S(p-i\partial)(-2m)\frac{\partial}{\partial p^\rho}b^\rho\gamma_5S(p-i\partial)\frac14(2p^\alpha-i\partial^\alpha)\gamma^\beta.
\en
By applying the momentum derivative on the $p$-functions to the right, $\Pi_b^{\mu\nu\alpha\beta} = \Pi_1^{\mu\nu\alpha\beta} + \Pi_2^{\mu\nu\alpha\beta} + \Pi_3^{\mu\nu\alpha\beta} + \Pi_4^{\mu\nu\alpha\beta}$ and  $\Pi_c^{\mu\nu\alpha\beta} = \Pi_5^{\mu\nu\alpha\beta} + \Pi_6^{\mu\nu\alpha\beta}$, with
\ba
\Pi_1^{\mu\nu\alpha\beta} &=& -\frac m8 \RM{tr} \int\frac{d^4p}{(2\pi)^4} S(p)\gamma_5S(p)\Slash{b}S(p)(2p^\mu-i\partial^\mu)\gamma^\nu S(p-i\partial)(2p^\alpha-i\partial^\alpha)\gamma^\beta, \\
\Pi_2^{\mu\nu\alpha\beta} &=& -\frac m8 \RM{tr} \int\frac{d^4p}{(2\pi)^4} S(p)\gamma_5S(p)(2b^\mu-i\partial^\mu)\gamma^\nu S(p-i\partial)(2p^\alpha-i\partial^\alpha)\gamma^\beta, \\
\Pi_3^{\mu\nu\alpha\beta} &=& -\frac m8 \RM{tr} \int\frac{d^4p}{(2\pi)^4} S(p)\gamma_5S(p)(2p^\mu-i\partial^\mu)\gamma^\nu S(p-i\partial)\Slash{b}S(p-i\partial)(2p^\alpha-i\partial^\alpha)\gamma^\beta, \\
\Pi_4^{\mu\nu\alpha\beta} &=& -\frac m8 \RM{tr} \int\frac{d^4p}{(2\pi)^4} S(p)\gamma_5S(p)(2p^\mu-i\partial^\mu)\gamma^\nu S(p-i\partial)(2b^\alpha-i\partial^\alpha)\gamma^\beta, \\
\Pi_5^{\mu\nu\alpha\beta} &=& -\frac m8 \RM{tr} \int\frac{d^4p}{(2\pi)^4} S(p)(2p^\mu-i\partial^\mu)\gamma^\nu S(p-i\partial)\gamma_5S(p-i\partial)\Slash{b}S(p-i\partial)(2p^\alpha-i\partial^\alpha)\gamma^\beta, \\
\Pi_6^{\mu\nu\alpha\beta} &=& -\frac m8 \RM{tr} \int\frac{d^4p}{(2\pi)^4} S(p)(2p^\mu-i\partial^\mu)\gamma^\nu S(p-i\partial)\gamma_5S(p-i\partial)(2b^\alpha-i\partial^\alpha)\gamma^\beta.
\ea 
The procedure used to evaluate the integrals (\ref{inta}), (\ref{intb}), and (\ref{intc}) follows the method developed in \cite{Ros,Adl} for the calculation of the axial anomaly in electrodynamics, which has also been employed in the calculations of the axial anomaly in gravity \cite{Kum} and in supergravity \cite{Nie}. Namely, we first write the most general tensor structure for $\Pi^{\mu\nu\alpha\beta}=\Pi_a^{\mu\nu\alpha\beta}+\frac12\Pi_b^{\mu\nu\alpha\beta}+\frac12\Pi_c^{\mu\nu\alpha\beta}$, in momentum space, as follows
\be
S_\RM{CS}[h] = i \int \frac{d^4k}{(2\pi)^4} \,h_{\mu\nu}(k)\,\Pi^{\mu\nu\alpha\beta}\,h_{\alpha\beta}(-k),
\en
where
\be
\Pi^{\mu\nu\alpha\beta} = \epsilon^{\mu\alpha\lambda\rho}b_\lambda k_\rho \left(Ag^{\nu\beta}+Bk^\nu k^\beta\right).
\en      
The form factor $A$ is at most logarithmically divergent and $B$ is finite. By employing the Feynman parametrization and evaluating the trace over the Dirac matrices, we obtain 
\ba
A &=& \frac{3im^2}{4}\int_0^1dx\int\frac{d^4p}{(2\pi)^4}\frac{(1-2x(1-x))(p^2-m^2)p^2+2x^2(1-x)^2 p^2k^2}{(p^2-M^2)^4} - \int\frac{d^4p}{(2\pi)^4} \frac{im^2}{(p^2-m^2)^2} \nonumber\\
B &=& \frac{3im^2}{4}\int_0^1dx\,(1-2x)^2\int\frac{d^4p}{(2\pi)^4}\frac{(1-2x)^2p^2-(1-2x(1-x))m^2+2x^2(1-x)^2 k^2}{(p^2-M^2)^4},
\ea
where $M^2=m^2-x(1-x)k^2$. Notice that $A$ and $-k^2B$ differ at most by a divergent constant which can be adjusted to be zero by a convenient renormalization prescription. This is automatically enforced if gauge invariance holds. Indeed, as the above action must be invariant under the gauge transformation $h_{\mu\nu}\to h_{\mu\nu}+k_\mu\Lambda_\nu+k_\nu\Lambda_\mu$, by imposing the transversality condition
\be
k_\mu\Pi^{\mu\nu\alpha\beta} = k_\nu\Pi^{\mu\nu\alpha\beta} = 0,
\en
we obtain that $A=-k^2B$, so that
\be
\Pi^{\mu\nu\alpha\beta} = \epsilon^{\mu\alpha\lambda\rho}b_\lambda k_\rho \left(-k^2g^{\nu\beta}+k^\nu k^\beta\right) B.
\en      
Finally, by integration over momenta and Feynman parameter, the form factor $B$ results in
\be
B = \frac{m^2}{16\pi^2(k^2)^{3/2}}\left[\sqrt{k^2}-\arctan\left(\frac{\sqrt{k^2}}{\sqrt{4m^2-k^2}}\right)\sqrt{4m^2-k^2}\right].
\en 
The leading contribution in the above expression as $k^2/m^2$ tends to zero is $B=1/192\pi^2$, which yields 
\be\label{CS2}
S_\RM{CS}[h] = \frac{1}{192\pi^2} \int d^4x \,b_\mu\epsilon^{\mu\nu\lambda\rho}h_{\nu\alpha}\partial_\lambda\left(\Box{h_\rho}^\alpha - \partial^\alpha\partial^\beta h_{\rho\beta}\right),
\en
in coordinates space. By expressing this Chern-Simons action in terms of the spin connection $\omega_{\mu ab}$, we have
\be\label{CS}
S_\RM{CS}[w] = \frac{1}{96\pi^2}\int d^4x
b_{\mu}\,\epsilon^{\mu\nu\lambda\rho}\left(
\omega_{\nu ab}\partial_\lambda{\omega_\rho}^{ba}-
\frac{2}{3}\omega_{\nu ab}{\omega_{\lambda}}^{bc}
{\omega_{\rho c}}^a\right).
\en
Therefore, by taking into account only the Chern-Simons contribution to the effective action (\ref{total}),
\ba
\Gamma_\RM{CS}[\omega]=C\int d^4x
b_{\mu}\,\epsilon^{\mu\nu\lambda\rho}\left(
\omega_{\nu ab}\partial_\lambda{\omega_\rho}^{ba}-
\frac{2}{3}\omega_{\nu ab}{\omega_{\lambda}}^{bc}
{\omega_{\rho c}}^a\right),
\ea
we can conclude that we obtain a gravitational Chern-Simons term with an undetermined constant $C$, generated by the undetermined additive term in the current. The ambiguity that we found probably can be interpreted as an implication of the gravitational anomalies \cite{Alv}. 

\section{Summary}

We have studied the problem of ambiguities in the linearized
gravity. It turns out that the mechanism of generating the
ambiguities due to possibility of modifying of the conserved
current by additive terms is applicable in the Lorentz-breaking
gravity as well as in the Lorentz-breaking electrodynamics \cite{Chung}, with the
ambiguity, depending on an arbitrary constant parameter.  In principle,
presence of the ambiguity in this theory is also confirmed by the fact
that the results obtained in \cite{Mar2} and in \cite{ptime} for the
gravitational Chern-Simons term are different. However, this mechanism
of arising of the ambiguity essentially differs from one
found earlier for the Lorentz-breaking QED which was based on use of
different regularization schemes for the formally divergent integrals
\cite{ourd4}. Presently the divergent integrals do not arise at
any steps of calculations. We also expect that a similar situation
will occur if we consider the complete, non-linearized expression for
the gravitational 
Chern-Simons term.

\vspace{5mm}

{\bf Acknowledgements.} This work was partially supported by CNPq and FAPESP.
The work by A. Yu. P. has been supported by CNPq-FAPESQ DCR program, CNPq project 350400/2005-9, and the work by T.~M. has been supported by FAPESP, project 06/06531-4.

\end{document}